\documentclass{article}

\usepackage{microtype}
\usepackage{graphicx}
\usepackage{subcaption}
\usepackage{booktabs} 
\usepackage{xspace}
\usepackage{tcolorbox}
\usepackage{pifont}
\usepackage{xcolor}
\usepackage{makecell}

\usepackage{hyperref}

\usepackage[preprint]{icml2026}

\usepackage{amsmath}
\usepackage{amssymb}
\usepackage{mathtools}
\usepackage{amsthm}
\usepackage{ulem}
\usepackage{hyperref}
\usepackage{url}
\usepackage{booktabs}
\usepackage{graphicx}
\usepackage{xspace}
\usepackage{caption}     
\usepackage{subcaption}
\usepackage{tcolorbox}
\tcbuselibrary{breakable}

\usepackage{booktabs}
\usepackage{multirow}

\usepackage[capitalize,noabbrev]{cleveref}

\theoremstyle{plain}

\theoremstyle{definition}

\theoremstyle{remark}

\usepackage[textsize=tiny]{todonotes}

\icmltitlerunning{\Name: Adversarial Claim Attacks against Search-Enabled LLM-based Fact-Checking Systems}

\newcommand{\Name}{\texttt{DECEIVE-AFC}\xspace}

\newenvironment{packeditemize}{
\begin{list}{$\bullet$}{
\setlength{\labelwidth}{6pt}
\setlength{\itemsep}{0pt}
\setlength{\leftmargin}{\labelwidth}
\addtolength{\leftmargin}{\labelsep}
\setlength{\parindent}{0pt}
\setlength{\listparindent}{\parindent}
\setlength{\parsep}{0pt}
\setlength{\topsep}{3pt}}}{\end{list}}

\newtcolorbox{mybox}[2][]{text width=0.95\linewidth,fontupper=\normalsize,
fonttitle=\bfseries\sffamily\scriptsize, colbacktitle=darkgrey,enhanced,
attach boxed title to top left={yshift=-2mm,xshift=3mm},
boxed title style={sharp corners},top=4pt,bottom=2pt,left=2pt,right=2pt,
  title=#2,colback=white}

\begin{document}

\twocolumn[
  \icmltitle{\Name: Adversarial Claim Attacks against \\Search-Enabled LLM-based Fact-Checking Systems}



  \icmlsetsymbol{equal}{*}

  \begin{icmlauthorlist}
      \icmlauthor{Haoran Ou}{ntu}
      \icmlauthor{Kangjie Chen}{ntu}
      \icmlauthor{Gelei Deng}{ntu}
      \icmlauthor{Hangcheng Liu}{ntu}
      \icmlauthor{Jie Zhang}{astar}
      \icmlauthor{Tianwei Zhang}{ntu}
      \icmlauthor{Kwok-Yan Lam}{ntu}
  \end{icmlauthorlist}

  \icmlaffiliation{ntu}{Nanyang Technological University, Singapore}
  \icmlaffiliation{astar}{CFAR, A*STAR, Singapore}

  \icmlcorrespondingauthor{Tianwei Zhang}{tianwei.zhang@ntu.edu.sg}

  \icmlkeywords{Machine Learning, ICML}

  \vskip 0.3in
]



\printAffiliationsAndNotice{}  

\begin{abstract}


Fact-checking systems with search-enabled large language models (LLMs) have shown strong potential for verifying claims by dynamically retrieving external evidence. However, the robustness of such systems against adversarial attack remains insufficiently understood. In this work, we study adversarial claim attacks against search-enabled LLM-based fact-checking systems under a realistic input-only threat model. We propose \Name, an agent-based adversarial attack framework that integrates novel claim-level attack strategies and adversarial claim validity evaluation principles. \Name systematically explores adversarial attack trajectories that disrupt search behavior, evidence retrieval, and LLM-based reasoning without relying on access to evidence sources or model internals. Extensive evaluations on benchmark datasets and real-world systems demonstrate that our attacks substantially degrade verification performance, reducing accuracy from 78.7\% to 53.7\%, and significantly outperform existing claim-based attack baselines with strong cross-system transferability.

\end{abstract}

\section{Introduction}
\label{sec:intro}

\begin{figure}
    \centering
    \includegraphics[width=1\linewidth]{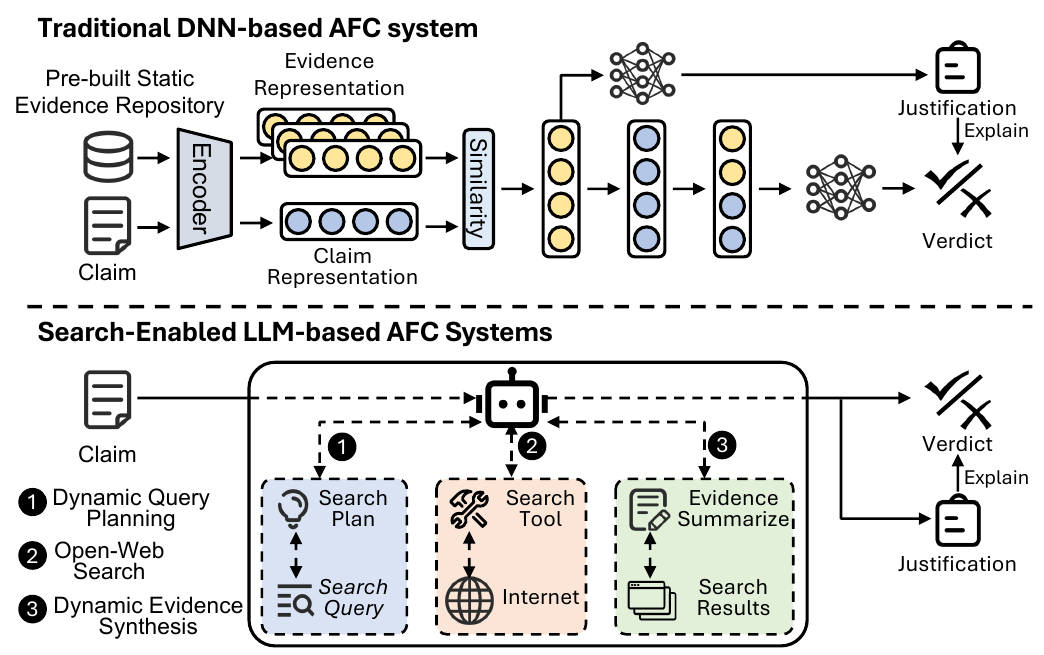}
    \caption{Comparison between traditional DNN-based AFC systems and search-enabled LLM-based AFC systems.}
    \label{fig:intro}
    \vspace{-15pt}
\end{figure}

The rapid spread of misinformation across online platforms poses serious societal risks, such as influencing public opinion and undermining trust in institutions. Although manual fact-checking conducted by professional journalists and domain experts is an effective countermeasure against misinformation, it is time-consuming and labor-intensive, making it difficult to keep pace with the rapidly growing volume of misinformation. To address this challenge, many automated fact-checking (AFC) methods have been proposed~\cite{akhtar2023multimodal,guo2022survey}, which can verify claims by automatically retrieving evidence and reasoning over it to produce a verdict and justification.

Currently, AFC methods can be broadly categorized into traditional Deep Neural Network (DNN)-based approaches and Large Language Model (LLM)-based systems. As illustrated in Figure~\ref{fig:intro}, traditional AFC systems typically retrieve evidence from static, pre-constructed repositories and rely on lightweight classification models to produce final verification decisions. The limited coverage and timeliness of such repositories constrain their ability to reliably handle newly emerging or rapidly evolving claims.
In contrast, LLMs’ strong reasoning and language understanding capabilities, along with their ability to perform web-based searches for up-to-date evidence, make them better suited for fact-checking tasks. Consequently, numerous LLM-based AFC methods have been proposed in recent years~\cite{llm_reasoning_1,llm_factcheck_1, llm_factcheck_2,search_llm_factcheck_1, search_llm_factcheck_2, search_llm_factcheck_3}. 

Although LLM-based AFC methods demonstrate strong misinformation detection capabilities, attackers can still exploit adversarial attack techniques to bypass the detectors and spread misinformation. According to~\cite{afc_attack_survey}, existing adversarial attacks targeting AFC methods can be divided into three types: adversarial claim attacks~\cite{facteval}, adversarial evidence attacks~\cite{ae_evidence}, and adversarial claim-evidence pair attacks~\cite{ae_claim_evidence}. 
Adversarial claim attacks aim to generate new claims or manipulate existing ones to mislead AFC systems. 
Compared to adversarial claim attacks, the other two attack paradigms typically require the ability to manipulate evidence sources or construct claim-evidence pairs within curated benchmarks~\cite{afc_attack_survey}, assumptions that are often unrealistic or inaccessible in real-world AFC systems with open web retrieval. In contrast, adversarial claim attacks only require the attacker to have control over the claim itself, without any additional impractical assumptions (e.g., falsify pre-built evidence corpus), which is more practical in the real world~\cite{afc_attack_survey}. Therefore, we focuses on adversarial claim attacks.
Existing adversarial claim attacks are largely designed for traditional DNN-based AFC pipelines, where evidence is retrieved from static corpora using simple mechanisms, or even bypassed entirely by directly providing golden evidence to the verifier. Such design choices impose inherent limitations when extending to search-enabled LLM-based fact-checking systems. These systems benefit from substantially stronger natural language understanding and reasoning capabilities, as well as more sophisticated claim interpretation and dynamic evidence retrieval strategies, making them far more robust to superficial character-level or word-level manipulations. As a result, \textit{\uline{existing adversarial claim attacks struggle to induce system-level verification failures}}, leaving a significant gap in the systematic study of attacks against search-enabled LLM-based AFC systems.



To address these limitations, we propose \Name, an agent-based claim-based attack framework for search-enabled LLM-based fact-checking systems. \Name consists of three tightly coupled components: a set of claim-level adversarial attack strategies, an attack validity evaluation mechanism, and an agent-based exploration framework.
Specifically, we first formalize a set of claim-level adversarial attack strategies tailored to search-enabled fact-checking systems, targeting different stages of the pipeline: search query formulation, evidence retrieval, and LLM-based reasoning. Meanwhile, a principled attack validity evaluation strategy is introduced, which constrains adversarial claim generation.
These strategies are designed to ensure that adversarial attacks remain meaningful by preserving the factual intent of the original claim, while subtly altering its linguistic or structural properties. Building on these strategies, adversarial claims can mislead the verification process without introducing explicit falsehoods.
Finally, we propose a novel agent-based framework to automatically and systematically explore adversarial attack trajectories, thereby generating effective and semantically valid adversarial claims.


Extensive evaluations demonstrate that \Name can consistently induce verification failures in search-enabled LLM-based fact-checking systems while preserving semantic validity. Moreover, the generated adversarial claims exhibit strong transferability across different real-world AFC systems, substantially outperforming existing claim-based attack baselines. It leads to a substantial degradation in target model performance, reducing verification accuracy from 78.7\% to 53.7\%. Further analyses reveal that these failures arise from systematic disruptions to evidence retrieval and downstream reasoning, rather than superficial perturbations, underscoring the effectiveness of our framework. 
Our contributions can be summarized as follows:
\begin{packeditemize}
    \item We propose an agent-based claim-level adversarial attack framework for search-enabled LLM-based fact-checking systems, which systematically explores adversarial attack trajectories under black-box settings.
    \item We formalize a set of claim-level adversarial attack strategies targeting different stages of the fact-checking pipeline, together with an attack validity evaluation mechanism that preserves label consistency and factual intent.
    \item We empirically demonstrate that existing claim-based attacks fail to generalize to search-enabled LLM-based AFC systems, and provide mitigation insights for improving system robustness against adversarial claim attacks.
\end{packeditemize}

\section{Related Work}
\label{sec:related-work}

\subsection{Automated Fact-Checking Systems}


Existing automated fact-checking (AFC) systems generally follow a standardized pipeline. Given an input claim, the system first retrieves relevant evidence, and then performs verdict prediction and justification generation based on the retrieved evidence~\cite{akhtar2023multimodal,guo2022survey,thorne2018automated}. While this high-level workflow is shared across different approaches, the underlying mechanisms and models used at each stage vary substantially. 
Accordingly, existing AFC systems can be broadly categorized into two classes (Figure~\ref{fig:intro}): DNN-based systems and search-enabled LLM-based systems. 

\noindent\textbf{DNN-based AFC systems.}
Traditional DNN-based AFC approaches rely on simple retrieval mechanisms and static evidence repositories. Specifically, prior works typically retrieve evidence from pre-constructed large document corpora by measuring semantic relevance between claims and candidate documents, using techniques such as lexical matching, embedding-based similarity, or learned retrieval models~\cite{AFC_retrieval_1, AFC_retrieval_2, AFC_retrieval_4}.
The retrieved evidence is then fed into neural verification models to predict claim veracity by jointly modeling claims and evidence~\cite{AFC_verify_1, AFC_verify_2, AFC_verify_4}. Despite their effectiveness, traditional AFC systems suffer from several limitations: they depend on supervised training on fixed datasets, limiting generalization and robustness, and rely on static evidence corpora, which constrain coverage and adaptability to emerging misinformation~\cite{guo2022survey, AFC_limitation_1}.

\noindent\textbf{LLM-based AFC systems.} Recent work has explored the use of LLMs for fact-checking, using techniques such as zero-shot, few-shot, and chain-of-thought reasoning to enhance verification. \cite{llm_factcheck_1, llm_factcheck_2, llm_factcheck_3, llm_factcheck_4}. 
Although LLMs demonstrate improved effectiveness and transferability compared to traditional task-specific models, these standalone LLM-based systems remain limited when relying only on the finite knowledge acquired during pretraining for verification. To overcome these limitations, subsequent work~\cite{search_llm_factcheck_1, search_llm_factcheck_2, search_llm_factcheck_3, hiss, lemma, defame} extends the workflows and has proposed augmenting LLM-based fact-checking systems with external search tools, enabling them to dynamically retrieve evidence from the open web. Particularly, these systems leverage LLMs to interpret input claims, generate search queries to search relevant information on the open web, and summarize the retrieved content as evidence for verification and justification.

\subsection{Adversarial Attacks against Fact-Checking}

As automated fact-checking systems are increasingly deployed, a growing body of work has investigated their vulnerability to adversarial attacks. From a pipeline perspective, prior studies identify two primary attack surfaces in AFC systems: the input claim and the retrieved evidence~\cite{afc_attack_survey}. Based on these surfaces, existing attacks can be broadly grouped into three categories.
\textit{Adversarial claim attacks} manipulate the input claim while preserving its factual intent, for example through paraphrasing, entity substitution, or temporal and numerical variations, with the goal of inducing incorrect verification outcomes~\cite{ae_claim_attack_1, ae_claim_attack_2, ae_claim_attack_3, facteval}. These attacks operate solely at the claim level and do not directly modify evidence sources.
\textit{Adversarial evidence attacks}~\cite{ae_evidence} target the evidence retrieval stage by injecting, modifying, or ranking misleading evidence to influence the verification result. Such attacks typically assume the ability to tamper with evidence repositories, curated knowledge bases, or retrieval indices.
\textit{Adversarial claim-evidence pair attacks}~\cite{ae_claim_evidence} jointly manipulate claims and their associated evidence, often by constructing challenging claim-evidence pairs within existing fact-checking benchmarks to expose model vulnerabilities.

The three types of attacks differ substantially in practical threat assumptions. Evidence-based and claim-evidence-based attacks often rely on strong control over evidence sources or benchmark construction, assumptions that are difficult to satisfy in real-world, search-enabled fact-checking systems that retrieve evidence dynamically from the open web. In contrast, adversarial claim attacks operate purely through user queries under black-box settings, making them significantly more practical. This motivates our focus on adversarial claim attacks against search-enabled LLM-based fact-checking systems.

\section{Threat Model}
\label{sec:threat}

We consider a practical AFC system $f$ that consists of evidence retrieval, verdict prediction, and justification generation. It can be formally represented as 
\begin{equation}
    f\big(c, E(c)\big) = (y_c, j),
\end{equation}
where $c$ is an input claim, $E(c)$ is a set of evidence retrieved to support or refute the claim $c$. $y_c \in \{0,1\}$ denotes a verification verdict, where $y_c = 0$ indicates that the claim is true and $y_c = 1$ otherwise, and $j$ is the corresponding generated justification. 

\begin{figure*}[t]
    \centering
    \includegraphics[width=0.95\linewidth]{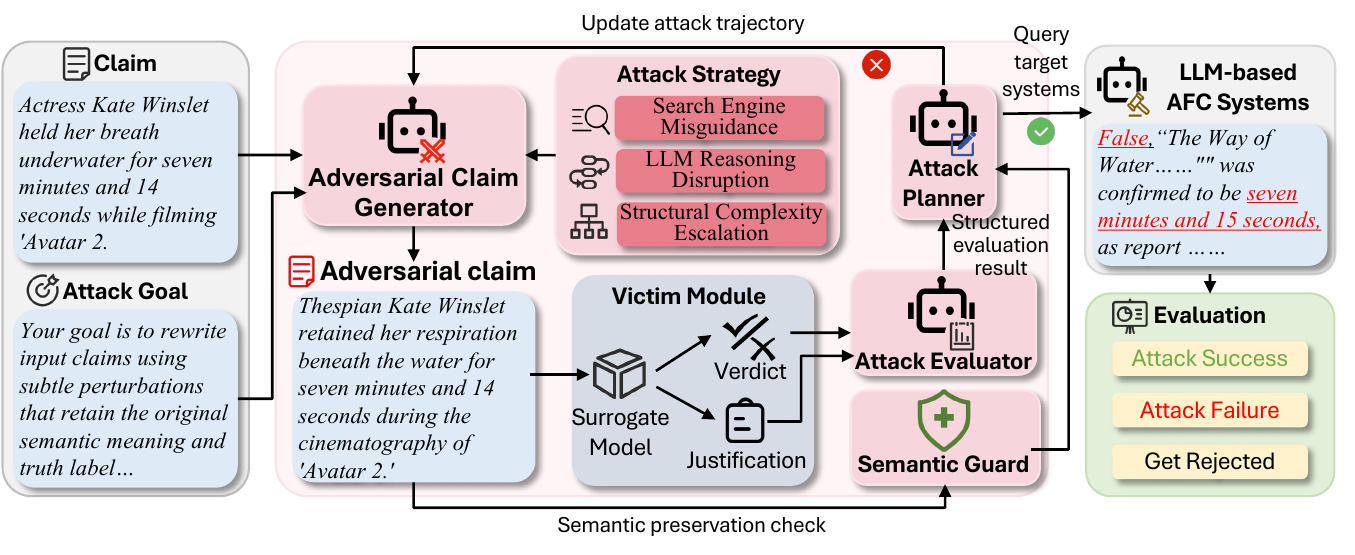}
    \caption{Overview of \Name.}
    \label{fig:framework}
\end{figure*}

\noindent\textbf{Attack goals.} 
We adopt a common and realistic threat scenario, adversarial claim attack, in which an attacker perturbs any $c$ into an adversarial claim $c'$ to mislead the automated fact-checking system.
A valid attack should simultaneously satisfy the following three conditions:
\begin{packeditemize}
    \item \textbf{Strong attack effectiveness.} The attack should reliably induce incorrect verification outcomes in the target AFC system, increasing the difficulty of combating misinformation and potentially amplifying the spread of misinformation in the real world.
    
    \item \textbf{Factual content preservation.} $c'$ should preserve the original semantics of $c$, i.e., $\mathrm{Sim}(c, c')\leq \tau$, where $\mathrm{Sim}(\cdot, \cdot)$ is a function that measures the semantic distance between the two inputs (a low value indicates a high similarity), and $\tau$ is a threshold for preserving semantic similarity. This constraint prevents semantic drift, ensuring that adversarial claims reflect the original factual content and that the verification remains meaningful.
    
    \item \textbf{High attack stealthiness.} $c'$ should induce relevant justifications, i.e., $\mathrm{Corr}(c', j_{c'}) \leq \gamma$, where $\gamma$ is also a pre-determined threshold. If the generated justification is completely meaningless or irrelevant, the attack can be easily detected through human inspection.
\end{packeditemize}  

\noindent\textbf{Attacker's capabilities.} To ensure the practicality of the attack, we assume the attacker can only access the inputs and outputs of the target AFC system, without any knowledge of the system internals (e.g., or evidence sources). This assumption is consistent with the typical deployment of commercial AFC systems. 
Nevertheless, the attacker can still generate $c'$ by freely querying a surrogate AFC system and leveraging transferability to launch the attack, which is a well-studied strategy in other domains.

\section{Methodology}
\label{sec:methodology}

To systematically investigate adversarial claim attacks against search-enabled LLM-based fact-checking systems, we present an agent-based adversarial evaluation framework (Figure~\ref{fig:framework}). We first introduce a set of claim-level adversarial attack strategies that define how input claims can be crafted to influence evidence retrieval and verification in search-enabled fact-checking pipelines (Section~\ref{subsec:attack_strategy}). Next, we specify attack validity evaluation strategy that constrains adversarial claim generation under realistic threat models (Section~\ref{subsec:attack_validity}).
Finally, based on these strategies, we design an agent-based pipeline to automatically discover effective adversarial claims through iterative refinement (Section~\ref{subsec:agent_pipeline}).

\subsection{Adversarial Claim Attack Strategy}
\label{subsec:attack_strategy}


Although attackers have no access to the internal models or retrieval mechanisms, the input claim serves as the sole interface through which the entire pipeline is activated. As a result, carefully crafted adversarial claims can implicitly influence multiple stages of the fact-checking process, shaping how evidence is retrieved and how the final verification decision is made. 
Motivated by this observation, we propose a set of adversarial claim attack strategies specifically designed for search-enabled fact-checking systems, aiming to induce incorrect verification outcomes under black-box settings. By targeting vulnerabilities at different stages of the fact-checking pipeline, we categorize adversarial claim attacks into three classes.

\noindent \textbf{1. Search engine misguidance.} Search-enabled fact-checking systems perform claim interpretation before retrieval and generate keywords or reformulated queries to search for relevant evidence. This strategy targets the online search component of the AFC system by subtly altering the lexical and structural properties of the input claim, including four sub-strategies: (1) \textit{low-frequency synonym introduction}, (2) \textit{non-standard entity referencing}, (3) \textit{redundant background information injection}, and (4) \textit{keyword dispersion}. 
Such manipulations implicitly affect query formulation and ranking behaviors, increasing the likelihood of retrieving irrelevant or low-quality evidence. Although these modifications preserve the overall factual intent of the claim, they reduce the prominence of key words for search. As a result, irrelevant, low-quality, or misleading evidence is more likely to be retrieved, and the subsequent verification is performed on an evidence set that appears plausible but is factually incorrect.

\noindent \textbf{2. LLM reasoning disruption.}
LLM-based fact-checking systems involve a reasoning process when organizing retrieved results into usable evidence and when performing final verification to produce labels and justifications. This strategy aims to mislead such reasoning processes to induce incorrect verification outcomes and justifications, including four sub-strategies: (1) \textit{injecting factually irrelevant but valid statements}, (2)  \textit{increasing syntactic complexity}, (3) \textit{introducing conditional or speculative phrasing}, and (4) \textit{employing double negation structures}. 
These manipulations distract the model’s attention, thus weakening its ability to associate evidence with the core factual claim correctly. These perturbations do not distort the factual correctness of the claim itself, but increase reasoning ambiguity and load, making the model more prone to erroneous conclusions or incoherent justifications.

\noindent \textbf{3. Structural complexity escalation.} This strategy targets the end-to-end verification pipeline by increasing the difficulty of claim verification. Specifically, it increases the structural complexity of the claim by transforming simple, single-hop factual assertions into multi-hop reasoning problems. This is achieved by two sub-strategies: (1) \textit{decomposing explicit entities into indirect references}, which obscures direct relations through intermediate concepts, and (2) \textit{rephrasing atomic facts as compound relational statements}. 
Such transformations force the fact-checking system to perform multi-step retrieval and reasoning to relocate the core factual content that needs to be verified. This escalation expands the retrieval and reasoning process from a single hop to multiple hops, making the pipeline more brittle: an error in any intermediate hop can break the chain, causing verification to fail even if other steps are correct. We detail the strategies description in Appendix~\ref{sec:detailed_attack_strategy}.


\subsection{Attack Validity Evaluation Strategy}
\label{subsec:attack_validity}

In this section, we describe how adversarial claim generation is constrained to ensure attack validity and how attack success is evaluated within our framework. Specifically, attack success is assessed at two stages of the pipeline: during the iterative refinement process and in the final evaluation stage. We adopt distinct evaluation criteria at each stage to reflect their different objectives and constraints.

\noindent \textbf{Attack validity evaluation during refinement.} In this process, attack success is determined by strict standard criteria using a fully automated pipeline. Specifically, we first require the verification verdict to be incorrect, ensuring that the attack induces a decision error.
We then use semantic similarity measures~\cite{semantic_similarity} to assess the consistency between the benign and adversarial claims, to prevent obvious semantic drift or topic changes that would invalidate the attack.
In addition, we employ Natural Language Inference (NLI)~\cite{NLI} to examine the logical relationship between benign and adversarial claims, ensuring that the adversarial claim preserves the original factual intent rather than altering it into a contradictory fact. Finally, we leverage an LLM to assess whether the generated justification is related to the adversarial claim, avoiding obvious topic drift. 

\noindent \textbf{Attack evaluation in the final stage.} In this stage, we adopt the same criteria but with softened thresholds. While a verdict flip remains mandatory, we relax the semantic similarity and NLI-based constraints: (1) the similarity threshold between benign and adversarial claims is lowered, and (2) the NLI requirement is weakened from strict bi-directional entailment to the absence of explicit contradiction.
Because misleading AFC systems often require subtle semantic shifts to induce incorrect fact-checking decisions, such shifts may reduce embedding-based similarity or NLI scores, even when the core factual intent is preserved.
For cases that fall below the strict thresholds in the refinement process, we conduct human validation to confirm that the adversarial claim does not alter the original factual intent.
In addition, human evaluators assess whether the adversarial claims remain readable for natural users. This evaluation strategy effectively ensures the validity of adversarial attacks, enabling to identify a broader set of successful attack cases.

\subsection{Agent-based Adversarial Refinement Framework}
\label{subsec:agent_pipeline}

Building upon the adversarial claim attack strategies (Section~\ref{subsec:attack_strategy}) as well as the attack validity and evaluation criteria (Section~\ref{subsec:attack_validity}), we design \Name, which can systematically discover effective adversarial claims under a realistic black-box scenario. \Name models the attack process as a closed-loop interaction among multiple components, each responsible for a distinct role in generation, evaluation, and refinement.

\noindent \textbf{Adversarial Claim Generator.} Conditioned on the selected attack strategy or feedback from previous iterations, the Generator is responsible for initializing candidate adversarial claims or refining existing ones. 
For each attack instance, the Generator performs multi-round refinement and retains context from previously generated claim variants and feedback within the same conversation, ensuring consistent and coherent modification of the claim.

\noindent \textbf{Victim Module.} To simulate the behavior of a realistic AFC system, we employ a surrogate fact-checking model as the victim model during refinement. This design avoids direct interaction with proprietary or rate-limited target systems, enabling efficient and repeated querying while remaining consistent with black-box settings. Given a benign claim or candidate adversarial claim, the Victim Module queries the surrogate model and returns the predicted verification verdict with corresponding justification.
Each interaction is performed independently, where no historical context or previous claim variants are retained across queries, preventing unintended contamination from earlier interactions and preserving fairness.

\noindent \textbf{Attack Evaluator.} Given the verification verdicts and justifications generated for both benign and adversarial claims, the Evaluator performs a structured and comprehensive analysis of how the AFC system’s behavior changes under adversarial perturbations across multiple interpretable dimensions, such as whether the verdict is flipped and the degree of shift in the generated justification.
Each evaluation is conducted independently without access to prior evaluation history, ensuring an unbiased assessment.

\noindent \textbf{Semantic Guard Module.}
In addition to the Evaluator, we employ a lightweight semantic guard to assess the validity of adversarial claims by checking for excessive semantic drift between benign and adversarial inputs.
This module ensures that adversarial claims do not degenerate into semantically unrelated or contradictory statements, retaining the original factual content. The specific semantic evaluation criteria follow those defined in Section~\ref{subsec:attack_validity}.

\noindent \textbf{Attack Planner.} 
The Attack Planner receives structured evaluation results from the Evaluator and semantic validity outcomes from the Semantic Guard, and dynamically adapts attack trajectories. Specifically, it first updates the current attack state based on inputs, indicating whether an adversarial attempt is successful and semantically valid. This attack state then determines subsequent actions, including whether to terminate the attack or continue refinement. If refinement continues, the Attack Planner further decides the direction of refinement, such as continuing optimization within the current strategy, switching to an alternative variant within the same strategy family, or transitioning to a different high-level attack strategy. This hierarchical control mechanism allows the framework to adaptively and effectively explore diverse attack paths for different failure modes observed during evaluation.
The refinement guidance is then passed to the Generator to guide subsequent claim generation.
\section{Experiments}
\label{sec:experiments}


\subsection{Setup}

\textbf{Target models.}
We evaluate \Name's performance on three representative search-enabled LLM-based fact-checking systems, which serve as the victim models in our experiments, including (1) HiSS~\cite{hiss}, (2) LEMMA~\cite{lemma}, and (3) DEFAME~\cite{defame}. These systems all build upon LLMs for fact-checking and dynamically retrieve external evidence from the open web. Detailed descriptions are provided in Appendix~\ref{subsec:target_models}.

\noindent \textbf{Baselines.}
We select four representative and effective claim perturbations proposed in FACTEVAL~\cite{facteval} as baselines, achieving high attack success rates across LLM-based fact verification systems. Specifically, the baseline attacks include \textit{LEET}, \textit{Homoglyph}, \textit{Character Swap}, and \textit{Phonetic} perturbations, covering visual obfuscation, character-level noise, and phonetic variation. Detailed descriptions of each baseline perturbation are provided in Appendix~\ref{subsec:baseline_attacks}.

\noindent \textbf{Dataset.}
Our experiments are conducted on the MOCHEG dataset~\cite{mocheg}, a benchmark designed to evaluate automated fact-checking (AFC) systems. Rather than directly generating claims, we adopt a claim manipulation setting, where \Name is used to transform benign claims collected from the real-world into adversarial claims.
In our evaluation, we conduct attacks on the test split of MOCHEG and exclude samples labeled as \textit{Not Enough Information} (NEI) to align with our threat model assumptions. After filtering, the evaluation set contains a total of 1,642 samples, including 817 positive and 825 negative claims.

\noindent \textbf{Metrics.} 
To evaluate the effectiveness of adversarial claim attacks against search-enabled LLM-based fact-checking systems, we report the following metric.

\begin{packeditemize}
    \item \textit{Attack Success Rate (ASR).} This is defined as the proportion of adversarial claims that successfully cause the target fact-checking system to produce an incorrect verification verdict under the validity constraints (Section~\ref{subsec:attack_validity}).
    
    \item \textit{ROUGE-1/2/L.} It measures lexical overlap between the two justifications, including unigram/bigram and longest common subsequence similarity~\cite{rouge}.

    \item \textit{MAUVE.} It reflects distribution-level similarity between the sets of generated justifications, measuring whether successful attacks induce a shift in the overall justification generation distribution~\cite{MAUVE}.
    
    \item \textit{SummaC (avg).} It assesses semantic agreement between the two justifications where,(higher indicates stronger semantic consistency; negative values indicate weaker agreement in our setting)~\cite{summac}.
\end{packeditemize}

In addition to ASR, we also report standard classification metrics, including accuracy, precision, recall, and F1 score to assess the overall verification performance of target models under benign and adversarial inputs.
To ensure that adversarial claims remain semantically consistent with their benign counterparts, we additionally report semantic similarity scores and Natural Language Inference (NLI) outcomes between benign and adversarial claims. These metrics are used to characterize the validity of adversarial attacks.

\noindent \textbf{Configuration.}
We evaluate \Name under a unified black-box setting, where different model components are assigned roles for adversarial claim generation, evaluation, and validation.
Both the Generator and Evaluator agents are instantiated using GPT-4o, while the surrogate fact-checking system is implemented with a search-enabled LLM that integrates external web retrieval.
Adversarial claims are iteratively refined under semantic preservation constraints with a fixed iteration budget.
Full implementation details, model configurations, and validation criteria are provided in Appendix~\ref{sec:appendix_config}.

\begin{table}[t]
    \centering
    \caption{Verification performance of the surrogate victim model under benign and adversarial claims.}
    \vspace{-3pt}
    \small
    \resizebox{\linewidth}{!}{
    \begin{tabular}{l c c c c c}
    \toprule
    \textbf{Setting} & \textbf{Acc} & \textbf{Prec} & \textbf{Rec} & \textbf{F1} & \textbf{ASR} $\uparrow$\\
    \midrule 
    Benign & 90.5\%   & 95.1\%   & 85.4\%   & 90.0\% & --\\
    \Name & 57.4\%   & 61.1\%   & 40.1\%   & 48.4\%  & 36.6\%\\
    \bottomrule
    \end{tabular}
    }
    \label{tab:surrogate_model_performance}
\end{table}

\begin{table*}[t]
    \centering
    \caption{Comparison of adversarial claim attack methods across different target fact-checking systems. ``Benign'' denotes verification performance on original (non-adversarial) claims. ASR is computed under the validity constraints described in Section~\ref{subsec:attack_validity}.}
    \resizebox{\linewidth}{!}{
    \begin{tabular}{l ccccc ccccc ccccc}
    \toprule
    \multirow{2}{*}{\textbf{Attack Method}}
    & \multicolumn{5}{c}{\textbf{HiSS}}
    & \multicolumn{5}{c}{\textbf{LEMMA}}
    & \multicolumn{5}{c}{\textbf{DEFAME}} \\
    \cmidrule(lr){2-6}\cmidrule(lr){7-11}\cmidrule(lr){12-16}
    & \textbf{Acc} & \textbf{Prec} & \textbf{Rec} & \textbf{F1} & \textbf{ASR} $\uparrow$
    & \textbf{Acc} & \textbf{Prec} & \textbf{Rec} & \textbf{F1} & \textbf{ASR} $\uparrow$
    & \textbf{Acc} & \textbf{Prec} & \textbf{Rec} & \textbf{F1} & \textbf{ASR} $\uparrow$ \\
    \midrule
    \textbf{Benign} 
    & 77.5\% & 80.8\% & 70.8\% & 75.4\% & -- 
    & 66.1\% & 72.8\% & 64.9\% & 62.3\% & -- 
    & 78.7\% & 86.9\% & 76.9\% & 81.1\% & -- \\
    \midrule
    \textbf{LEET} 
    & -- & -- & -- & -- & -- 
    & 55.2\% & 69.2\% & 55.1\% & 44.8\% & 12.1\% 
    & 70.6\% & 70.9\% & 70.6\% & 70.5 & 8.9\% \\
    \textbf{Homoglyph} 
    & 69.1\% & 67.1\% & 67.2\% & 67.1\% & 8.5\% 
    & 59.7\% & 67.8\% & 54.9\% & 45.2 & 6.7\% 
    & 71.9\% & 71.2\% & 73.4\% & 72.3\% & 7.1\% \\
    \textbf{Character Swap}
    & 70.1\% & 70.5\% & 68.5\% & 69.5\% & 7.6\% 
    & 59.1\% & 58.7\% & 60.5\% & 59.5\% & 7.2\%
    & 73.1\% & 72.4\% & 74.1\% & 73.2\% & 5.8\% \\
    \textbf{Phonetic}
    & 73.4\% & 84.9\% & 54.2\% & 66.2\% & 4.3\% 
    & 57.5\% & 69.9\% & 57.5\% & 49.6\% & 8.8\% 
    & 71.9\% & 71.9\% & 71.8\% & 71.8\% & 6.9\% \\
    \midrule
    \textbf{\Name} 
    & 61.9\% & 62.2\% & 59.9\% & 61.1\% & \textbf{19.5\%}
    & 51.1\% & 50.6\% & 79.6\% & 61.9 & \textbf{18.8\%}
    & 53.7\% & 55.5\% & 58.4\% & 50.3\% & \textbf{31.4\%} \\
    \bottomrule
    \end{tabular}}
    \label{tab:baseline_vs_targets_full}
\end{table*}

\subsection{Main Results}

\noindent \textbf{Surrogate model performance.}
Table~\ref{tab:surrogate_model_performance} presents the surrogate model's verification performance under benign and adversarial claims.
Under benign inputs, the model achieves high verification accuracy. However, when evaluated on adversarial claims generated by \Name, its performance drops, with accuracy decreasing from 90.5\% to 57.4\% and F1 score dropping from 90.0\% to 48.4\%.
This significant degradation demonstrates that our attack method is highly effective at inducing incorrect verification outcomes against a strong search-enabled LLM-based fact-checking model, achieving a ASR of 36.6\%.

The high performance observed under benign settings is expected and has been noted in prior work~\cite{holmes}.
Specifically, GPT-4o-search-preview can often retrieve authoritative fact-checking articles that directly contain the original claims and their associated verdicts. For this reason, such models are not intended to serve as standalone fact-checking systems in realistic deployments. Nevertheless, their strong performance, combined with ease of deployment, fast inference, and moderate cost, makes them well-suited as surrogate models for large-scale adversarial evaluation under black-box settings.

\begin{figure}
    \centering
    \includegraphics[width=1\linewidth]{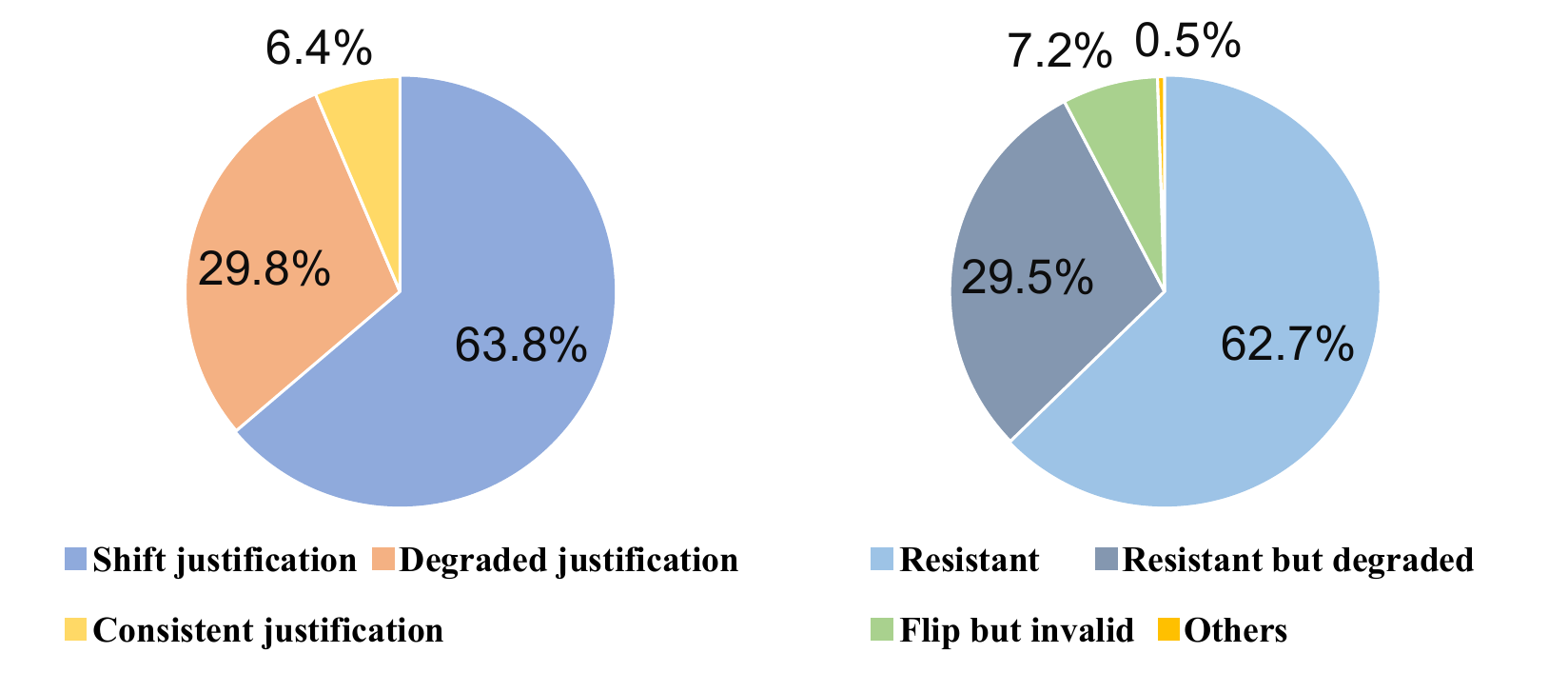}
    \vspace{-3pt}
    \caption{Outcome analysis of successful (left) and failed (right) adversarial optimizations.}
    \label{fig:refinement_analysis}
    \vspace{-10pt}
\end{figure}

\noindent \textbf{Refinement cases analysis.} Figure~\ref{fig:refinement_analysis} demonstrates the outcome distributions of successful and failed adversarial cases during optimizations. Successful cases predominantly arise from shifts or degradations in justification and evidence reasoning. In contrast, failures are largely due to strong model resistance or degraded reasoning without verdict flips, with a small fraction invalidated by semantic constraints. This suggests that effective attacks rely on manipulating justification and evidence interpretation.

\noindent \textbf{Target search-enabled LLM-based systems performance.} Table~\ref{tab:baseline_vs_targets_full} compares the effectiveness of different adversarial claim attacks across three target fact-checking systems. Overall, our method achieves the highest ASR results in all cases, demonstrating its ability to reliably generate semantically valid adversarial claims and strong transferability. Traditional methods lead to limited performance degradation, indicating that they are largely insufficient to induce valid adversarial claims. We note that under the LEET attack, HiSS produces a large number of refusal predictions, as the extensive spelling and typographical distortions render the claims unclear and incoherent. Such heavily distorted inputs are also difficult for humans to read, suggesting limited practical relevance in real-world fact-checking scenarios. As a result, LEET results on HiSS are not reported.

\noindent \textbf{Justification analysis.}
We compare the similarity between justifications generated for successful/failed adversarial claims and their original benign counterparts, relatively (Table~\ref{tab:justif_similarity}). Overall, justifications associated with attack failures exhibit substantially higher similarity across all metrics, indicating that unsuccessful attacks tend to preserve reasoning patterns closer to the benign case. In contrast, successful attacks produce justifications that are significantly less similar, reflecting pronounced shifts in evidence usage and reasoning structure. This gap suggests that effective adversarial claims primarily succeed by inducing meaningful deviations in the justification process, highlighting our proposed attack strategy is a key factor underlying successful attacks.

\begin{table}[t]
    \centering
    \caption{Justification similarity between success/failure claims and their original benign counterparts.}
    \label{tab:justif_similarity}
    \resizebox{\linewidth}{!}{
    \begin{tabular}{lccccc}
    \toprule
    Setting & ROUGE-1 & ROUGE-2 & ROUGE-L & MAUVE & SummaC \\
    \midrule
    Attack Success & 0.470 & 0.257 & 0.324 & 0.385 & -0.389 \\
    Attack Failure & \textbf{0.617} & \textbf{0.423} & \textbf{0.465} & \textbf{0.945} & \textbf{-0.193} \\
    \bottomrule
    \end{tabular}}
\end{table}

\subsection{Ablation Study}

To better understand the contributions of individual components in \Name, we conduct an ablation study focusing on two key factors that influence the performance of our approach: (1) the semantic guard module, and (2) the maximum number of refinement rounds. 

\begin{figure}[t]
    \centering
    \includegraphics[width=1\linewidth]{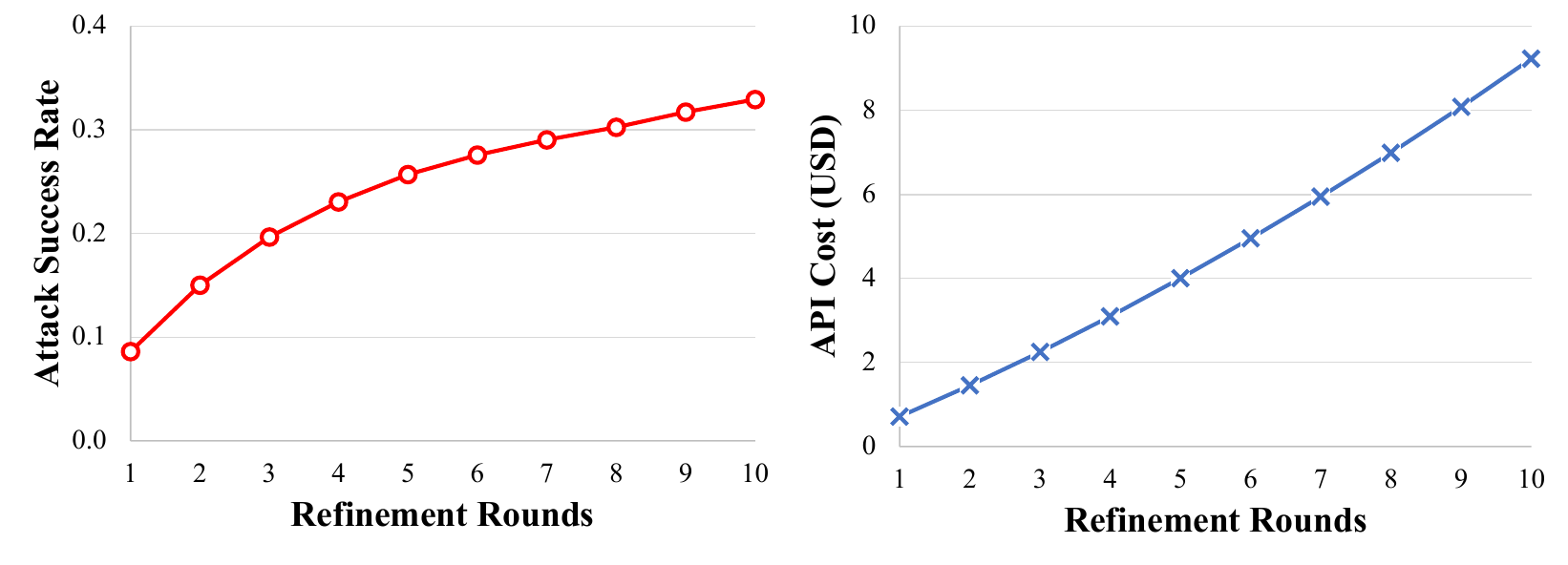}
    \caption{Impact of the maximum number of refinement rounds on ASR (left) and cost(right).}    
    \label{fig:ablation_refinement_rounds}
    \vspace{-10pt}
\end{figure}

\noindent \textbf{Refinement rounds.}
We investigate the effect of varying the number of refinement rounds from 1 to 10, keeping all other components fixed. We record the ASR and the refinement cost (100 samples per) as shown in Figure~\ref{fig:ablation_refinement_rounds}. Our results reveal a clear trade-off between optimization effectiveness and efficiency. Increasing the number of refinement rounds improves attack success rates, but with diminishing returns beyond a certain threshold. Excessive refinement can lead to higher computational costs and occasional over-optimization, where improvements become marginal. These findings suggest that a moderate number of refinement rounds strikes a balance between attack performance and efficiency.

\noindent \textbf{Semantic guard module}.
We examine the role of the semantic guard module by comparing our full framework with a variant in which the semantic guard is disabled. In this setting, adversarial claims are refined without enforcing semantic consistency between benign and adversarial claims, while the same semantic validity constraints are still applied at the final evaluation stage. We measure both the probability of label flipping and the overall attack success rate. The results (Table~\ref{tab:semantic_guard_ablation}) show that disabling the semantic guard slightly increases the frequency of label flips, indicating that unconstrained refinement can more easily induce incorrect intermediate verification outcomes. However, despite this higher flip rate, the overall attack success rate decreases, as many flipped cases fail to satisfy semantic validity constraints during final evaluation. Therefore, removing the semantic guard leads to less stable optimization behavior. Although label flipping becomes marginally more frequent, the number of valid successful attacks is significantly reduced, since a large fraction of flips are semantically invalid. These findings indicate that the semantic guard plays a crucial role in balancing attack effectiveness with semantic validity, acting as a guiding mechanism rather than a filter.

\section{Discussion about Potential Mitigation}
\label{sec:mitigation}

Our experimental results demonstrate that adversarial claim attack can successfully flip the final verification verdict. Meanwhile, generated justifications are seemingly relevant and coherent to the verdict, increasing the difficulty of identification. Incorrect decisions are no longer accompanied by obviously flawed reasoning, resulting in more severe and insidious risks. Even the search-enabled, LLM-based AFC systems, which are commonly regarded as more robust due to their multi-stage verification pipelines, remain susceptible to such attacks.
To mitigate these risks and reduce their potential harm, we propose two potential mitigation strategies, aiming at strengthening the robustness and safety of search-enabled LLM-based AFC systems.
The first is \textit{verification model fine-tuning}.
Most existing fact-checking systems rely on general base LLMs, lacking safety alignment to handle adversarially crafted claims. By fine-tuning base LLMs on adversarial datasets consisting of paired benign claims and their corresponding adversarial variants, the model can learn common adversarial perturbation patterns, thereby improving its ability to correctly interpret their underlying semantics.
The second is \textit{interpretation and retrieval validation.}
On the other hand, our results underscore the importance of strengthening the fact-checking pipeline itself. In search-enabled, LLM-based AFC systems, the verification process typically involves interpreting the input claim, generating search queries, and verifying the claim based on retrieved evidence. Adversarial claim manipulations can disrupt these stages, leading to the use of misleading, noisy, or low-credibility sources, even when the LLM is reasoning correctly. Introducing adversarial-aware correction or validation mechanisms at these intermediate stages can mitigate such risks by identifying and correcting erroneous queries and unreliable evidence.

\begin{table}[t]
    \centering
    \caption{Ablation study on the effect of the semantic guard.}
    \small
    \resizebox{\linewidth}{!}{
    \begin{tabular}{lcc}
        \toprule
        \textbf{Setting} 
        & \textbf{Label Flip Rate} 
        & \textbf{Attack Success Rate} \\
        \midrule
        Full framework 
        & 42.1\% 
        & 36.6\% \\
        w/o semantic guard 
        & 43.3\% 
        & 23.1\% \\
        \bottomrule
    \end{tabular}
    }
    \label{tab:semantic_guard_ablation}
\end{table}

\section{Conclusion}
\label{sec:conclusion}

In this paper, we conduct a systematic study of adversarial claim attacks against search-enabled LLM-based fact-checking systems. To address the limitations of prior works, we propose \Name, an agent-based adversarial claim attack framework that integrates novel attack strategies and attack validity constraints. 
These two components ensure our framework to launch effective attacks that degrade model performance, while remain semantically meaningful, retaining the original factual content.
Extensive experiments on multiple real-world search-enabled LLM-based AFC systems demonstrate that \Name consistently outperforms strong baseline attacks and exhibits robust cross-system transferability.
Our findings reveal that, despite their advanced reasoning capabilities, search-enabled LLM-based fact-checking systems remain vulnerable to carefully crafted adversarial claims that indirectly manipulate retrieval and verification processes. 
In addition, we discuss potential mitigation strategies aimed at refining existing approaches and facilitating the design of more robust, secure, and trustworthy fact-checking frameworks.







\clearpage
\section*{Impact Statement}
\label{sec:impact_statement}

This work aims to investigate the robustness and security of automated fact-checking systems, particularly search-enabled LLM-based frameworks. The primary goal of this research is to identify vulnerabilities in existing systems to assist in the design of more reliable, robust, and trustworthy fact-checking systems.
While adversarial attack techniques could potentially be misused to sabotage deployed systems, all adversarial examples are evaluated solely in controlled experimental environments.
By exposing failure modes in current fact-checking systems, this research aims to help the development of improved defenses and mitigation strategies, thereby helping reduce the spread of misinformation and enhance public trust in automated verification systems.




\nocite{langley00}

\bibliography{sample-base}
\bibliographystyle{icml2026}

\newpage
\appendix
\onecolumn
\appendix

\section*{Appendix}
\label{sec:appendix}

\section{Detailed Attack Strategy}
\label{sec:detailed_attack_strategy}

In this section, we provide detailed descriptions of the adversarial claim attack strategies used in our framework.

\noindent \textbf{1. Search Engine Misguidance.}
These strategies aim to indirectly influence the evidence retrieval stage by modifying lexical features of the claim, including four sub-categories:

\begin{packeditemize}

    \item \textit{Low-frequency synonym introduction.}
    This strategy replaces common nouns, verbs, or descriptors in the claim with low-frequency, formal, or uncommon synonyms. While semantic equivalence is preserved, the modified wording reduces alignment with typical search queries, thereby weakening keyword-based retrieval performance.

    \item \textit{Non-standard entity referencing.}
    Instead of explicitly mentioning named entities, this strategy refers to them indirectly through descriptive phrases, metonymy, or role-based identifiers. Although the referenced entity remains unambiguous to humans, the absence of canonical entity names can significantly affect search result relevance.

    \item \textit{Redundant background information injection.}
    This strategy inserts additional but factually irrelevant background details into the claim. These perturbations do not alter the claim’s truth value but introduce distracting keywords that may bias retrieval toward tangential or less informative evidence sources.

    \item \textit{Keyword dispersion.}
    This strategy restructures the claim by spreading core keywords across multiple clauses or diluting them with connective phrases. By lowering keyword concentration, this strategy weakens the search engine’s ability to correctly identify and prioritize relevant evidence.
    
\end{packeditemize}

\noindent \textbf{2. LLM Reasoning Disruption.}
These strategies target the verification stage by increasing the cognitive and reasoning complexity of the claim, including four sub-categories:

\begin{packeditemize}

    \item \textit{Injecting factually irrelevant but valid statements.}
    This strategy adds neutral, factually correct statements before or after the core claim. Although unrelated to the verification target, such statements can distract the model’s attention and interfere with its ability to focus on the relevant factual content.

    \item \textit{Increasing syntactic complexity.}
    The claim is rewritten using embedded clauses, nested sentence structures, or parenthetical expressions. These syntactic transformations preserve meaning but increase parsing difficulty, which can impair reasoning accuracy in complex verification scenarios.

    \item \textit{Introducing conditional or speculative phrasing.}
    This strategy embeds the claim within conditional or hypothetical constructions that do not alter its truth value. Such framing can obscure the model’s interpretation of factual assertions, leading to uncertainty or misclassification.

    \item \textit{Employing double negation structures.}
    The claim is rewritten using double negation or logically redundant constructions. Although logically equivalent to the original statement, these formulations increase reasoning load and can confuse entailment-based verification mechanisms.

\end{packeditemize}

\noindent \textbf{3. Structural Complexity Escalation.}
Structural complexity escalation strategies target multi-hop reasoning by transforming direct factual statements into more complex relational structures, including two sub-categories:

\begin{packeditemize}

    \item \textit{Decomposing explicit entities into indirect references.}
    This strategy replaces direct entity mentions with indirect descriptions involving intermediate concepts or attributes. As a result, the system must infer entity identity through multiple reasoning steps rather than relying on explicit mentions.

    \item \textit{Rephrasing atomic facts as compound relational statements.}
    Simple factual assertions are reformulated as compound statements involving multiple relationships or dependencies. While the underlying fact remains unchanged, verification now requires aggregating information across multiple relational hops.

\end{packeditemize}

\section{Detailed Experiment Settings}
\subsection{Target Models.}
\label{subsec:target_models}

This section provides brief descriptions of the target fact-checking systems evaluated in our experiments.

\noindent \textbf{HiSS.} It is an LLM-based fact-checking approach that leverages in-context learning for claim verification. Its core design introduces a Hierarchical Step-by-Step prompting strategy, which decomposes an input claim into multiple subclaims and verifies each subclaim through a sequence of question-answering steps.

\noindent \textbf{LEMMA.} It is a misinformation detection framework built upon large language models. The method enhances LLM-based reasoning by explicitly augmenting it with external knowledge. Given a claim, LEMMA generates multiple search queries to retrieve relevant external evidence. This retrieved knowledge is then incorporated into the LLM’s reasoning process to support verification decisions. 

\noindent \textbf{DEFAME.} It is a modular, zero-shot, search-enabled fact-checking system for open-domain claim verification. It adopts a multi-stage pipeline that dynamically selects tools and search depth to retrieve evidence from the open web. DEFAME performs end-to-end verification by jointly reasoning over claims, retrieved evidence, and produces verification reports.

\subsection{Baselines}
\label{subsec:baseline_attacks}

This appendix provides detailed descriptions of the baseline claim perturbations adopted from FACTEVAL~\cite{facteval}.

\noindent\textbf{LEET Perturbation.}
LEET perturbation replaces alphabetic characters with visually similar numbers or symbols (e.g., \textit{A} $\rightarrow$ \textit{4}, \textit{E} $\rightarrow$ \textit{3}). This writing style is commonly observed in online communities and adversarial settings. The perturbation primarily targets tokenization and lexical matching, while remaining easily interpretable by human readers.

\noindent \textbf{Homoglyph Perturbation.}
Homoglyph perturbation substitutes characters with visually confusable Unicode counterparts (e.g., \textit{n} $\rightarrow$ \textit{\~n}), based on standardized Unicode confusable character mappings. This attack introduces subtle visual ambiguity at the character level and has been shown to be particularly effective in misleading fact verification models without altering the perceived meaning of the claim.

\noindent \textbf{Character Swap Perturbation.}
Character Swap perturbation randomly swaps adjacent characters within words, simulating common typographical errors frequently observed in real-world text inputs. This perturbation introduces lightweight spelling noise while largely preserving word readability and sentence-level semantics.

\noindent \textbf{Phonetic Perturbation.}
Phonetic perturbation modifies words according to human-written phonetic variations, reflecting spelling patterns influenced by pronunciation differences, especially among non-native speakers. A fixed perturbation budget is applied to alter a subset of words in the claim, introducing realistic orthographic variations while maintaining semantic consistency.

\section{Implementation and Configuration Details}
\label{sec:appendix_config}

\subsection{Model Instantiation}
The Generator Agent and Evaluator Agent are both instantiated using GPT-4o.
The surrogate fact-checking system is implemented using GPT-4o-search-preview, which combines LLM-based reasoning with external web search.

\subsection{Semantic Guard and Validation Criteria}
Semantic similarity between the original and adversarial claims is computed using Sentence-BERT (all-mpnet-base-v2).
Natural Language Inference consistency is assessed using DeBERTa-large-MNLI.
During the refinement stage, adversarial claims are required to satisfy a semantic similarity threshold of 0.85 and bidirectional entailment under NLI.
For final evaluation, the similarity threshold is relaxed to 0.7, and the NLI constraint is weakened to the absence of explicit contradiction.
For cases that only satisfy the relaxed criteria, human validation is conducted to ensure preservation of factual intent.

\subsection{Refinement Budget and Strategy Space}
Each attack instance is refined for at most 10 iterations.
The attack strategy space consists of all combinations of strategy categories defined in Section~\ref{subsec:attack_strategy}.
Only adversarial claims that satisfy the semantic preservation criteria are used to replace the original claims in the constructed evaluation dataset.


\end{document}